\documentclass[12pt]{article}
\usepackage{graphicx}
\usepackage{amsfonts}
\usepackage{amscd}
\usepackage{latexsym}
\usepackage{epsf}

\newcommand{\be}{\begin{equation}}
\newcommand{\ee}{\end{equation}}
\newcommand{\bea}{\begin{eqnarray}}
\newcommand{\eea}{\end{eqnarray}}

\newcommand{\fr}{\frac}

\newcommand{\la}{ \lambda }

\newcommand{\ra}{\rightarrow}

\usepackage[latin9]{inputenc}
\usepackage{textcomp}
\usepackage{amsmath}
\usepackage{multirow}
\usepackage{color}
 
\newcommand {\si} {\sigma}

\newcommand {\bT} {\bar{T}} 
\begin{document}
\title{\bf 
 } 
\title{\bf
Effective Volumetric Factor for Inhomogeneous Temperature 
in Volumetric Measurements of Gas Absorption and Desorption}
\author{
{\sc I.~V.~Drozdov}\thanks{Corresponding author, e-mail: drosdow@uni-koblenz.de},
  \\}
\maketitle
\begin{center} 
\small Institute of Energy Research, 
\small  Forschungszentrum J\"ulich \\
\end{center} 

\begin{abstract}
Accounting of {\it effective volumetric factor} for the measurement at inhomogeneous temperature allows to the simplify the measurement process avoiding a pre-calibration with a noble gas.
 The volumetric correction in form of the volumetric factor is calculated analytically by  modeling of the temperature profile in the volumetric apparatus, and provides a quite 
 agreement with the measured results.
 The calculation can be applied for each volumetric measurement 
with an inhomogeneous temperature distribution. 
\end{abstract}


\section{Introduction}

The measurements of the gas absorption and desorption are mostly performed in a volumetric equipment.
A precise control of the gas amount at the inhomogeneous temperature distribution 
 is an essential issue for correct evaluation of experimental data.

 Volumetric Sievert's apparatuses  are widely used for experiments on gas absorption and desorption.
 They are relatively low cost equipment compared to more expensive gravimetrical equipment. 

A volumetric apparatus like represented in \cite{patent} consists conventionaly of a volumetric pool 
and a sample holder with the sorption material. During the adsorption/desorption,
 the sample is kept at the certain temperature. This temperature is not the same,
 as in the volumetric pool. 

In the exceptional case, when the temperature $T$ of the sorption gas is homogeneous distributed within the measurement setup,
 the normal state equation is used.
\be
\nu = \fr{p}{R} (V/T),
\label{state_eq}
\ee
to obtain a molar amount $\nu$ of the gas at the pressure $p$, $V$-total volume,
$R$-molar gas constant. 

If the gas is stored in several volumes $V_1, V_2,..., V_n$ at different
 temperatures $T_1, T_2, ... T_n$, the total molar amount is calculated as:
\be
\nu = \fr{p}{R} \{V/T\}; \mbox{ where } \{V/T\}=V_1/T_1+V_2/T_2+ ... + V_n/T_n,
\ee
and is defined as the {\it effective volumetric factor}, that should be used instead of the
 ratio V/T in (\ref{state_eq}).
In case, when the sample holder is heated/cooled, the gas temperature is distributed 
inhomogeneous. To establish the resulting volumetric factor ${V/T}$, some empirical ways are possible.
 One can use a non-adsorptive gas before the main measurement starts.
 In our measurements, the volumetric setup with the static inhomogeneous 
temperature distribution has been filled with helium. 
After two measurements of pressure in states with closed and open sample holder respectively,
the equipment has been completely evacuated and the adsorptive
 measurement gas (hydrogen) was introduced to the system.

 This simple and convenient method has certain shortcomings. There are some of these:

1. The pre-measurement including the following exhausting is significantly time consuming,
compared to the duration of the main measurement.

2. This procedure could not be applied micro- and nano-porous materials,
due to possible adsorption in porosities also for helium gas. 

 
3. The method is not applied for the measurements with variable temperature. 

 In the present work we have modelled the temperature profile in the volumetric apparatus 
 with respect to the temperature measured in several points. 
The expected {\it volumetric factor} for this  temperature distribution is calculated analytically.
%
 The experimental verification shows an exact agreement  with the measured result.

\section{Volumetric Principle}

 The calculations and verification were performed for the volumetric apparatus
BELSORP-HP, BEL Inc.Japan. The construction and principle of measurement
 is briefly described, as follows:
Before the main measurement, the active sorption volume (`dead volume')  
 $V_d=V_p+V_H$ should be determined using the helium gas.
 The proper volume of the sample holder $V_H$ is calculated from its geometry.
\begin{figure}  
 \begin{minipage}[b]{5cm}
\includegraphics[scale=0.40]{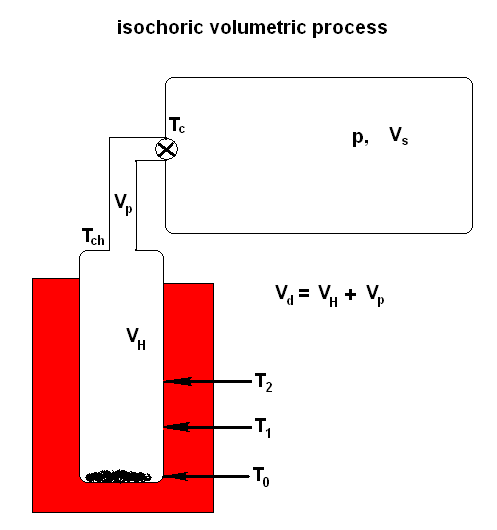}\end{minipage}
\hspace*{2cm}
\begin{minipage}[b]{6cm}
$V_H$ - the volume of the holder with the sample and hot gas in the thermostat at $T_0$\\
$V_p$ - volume of the pipe channel containing the gas with the descendent temperature from
 $T_{ch}$ down to $T_c$\\
$V_s$ - the system volume behind the ventil, with the cold gas of temperature $T_c$\\  
$T_1, T_2, T_{ch} $ the temperature measured in control points.\\
$T_c, T_0$ are set by the experiment conditions.\\ 
\end{minipage}
\caption{Structural scheme of the volumetric apparatus.} 
\label{scheme}
\end{figure}
  The molar amount of adsorbed/desorbed gas is evaluated from the difference of pressures 
 in initial $n$ and final $n+1$ states: 
\be
\nu_{ads/des}^{n\ra n+1} = \fr{(p_n-p_{n+1}) \{V/T\}    }{R}
\ee

The initial state $n$ corresponds to closed connection between the holder $V_p, V_H$ and the
volumetric pool $V_s$, whereat the pressure in the $V_H$, and the gas in $V_H+V_p$ remains
at the pressure of the previous step $p_{n-1}$.

 Each sorption step starts by opening of this connection.
 The pressure inside the setup (chamber) balances and the sorption begins.
 The current step is stopped in two cases,namely:\\

1) if no alteration of pressure is observed, or otherwise\\

2) after the certain time.

 The pressure reached in this step corresponds to the final state $p_{n+1}$.
 The measurement principle and procedure is described in details elsewhere \cite{BELmanual}.

\section{Accounting of the Temperature Inhomogeneity}
\label{inhomogeneity}

\subsection{Basic assumptions on the Temperature Distribution }

 The sample holder is typically designed as a pipe. All other gas connectors and conductors
 of the volumetric apparatus are pipes as well. 

 Hence, the following suggestion is commonly applied: the inhomogeneous temperature distribution
 is modelled for gas containing parts with translational symmetry. 
Let $z$ be a length coordinate along a pipe.  
The loss rate of the heat energy [J/s] in the surrounding air (thermic bath) of the 
temperature $\bar{T}$ with an empirical heat transport coefficient $\kappa$ per segment 
$\Delta\si$ of surface is supposed to be:
\be
\Delta \dot{Q}=\kappa [ T(z)- \bT]\Delta\si,
\ee
and the infinitesimal decrease of temperature at the point $z$ caused thereby is:
\be
\Delta T(z)=- \la C \Delta \dot{Q}=-\la C\kappa [ T(z)- \bT]\Delta\si
\ee

Assuming a translational symmetry of the gas container (pipe), $\Delta \si= S \Delta z$, $S$-
perimeter of the cross-section, $\Delta z$-infinitesimal part of length, and changing to differential
 values, one obtains:
\be
dT (z) =-\la S C \kappa [ T(z)- \bT ] dz := -\tau[ T(z)- \bT] dz.
\ee
An immediate integration with the initial condition $T(z)|_{z=z_0}=T_0$ provides an exponential decrease
of the temperature along the pipe container / channel, characterized by its cumulative parameter $\tau$
\be
T(z)=(T_0-\bT)e^{-\tau(z-z_0)}+\bT,
\ee
the simple ''shifted exponential descent'', that will be used further 
as an ansatz of the temperature distribution along the sample holder (also treated as a gas temperature).\\

The parameters $\tau, z_0$ of this distribution 
are provided by measurements of $T$ in two points $z_1,\ z_2=2z_1$ to be
\be
\tau=\fr{1}{z_1-z_0} \ln \fr{T_0-\bT}{T_1-\bT}=\fr{1}{2z_1-z_0} \ln \fr{T_0-\bT}{T_2-\bT},
\ee

\be
   z_0 = z_1 \left[1-\fr{1}{ \ln \fr{T_0-\bT}{T_2-\bT}/\ln \fr{T_0-\bT}{T_1-\bT}-1} \right] 
\ee
\\

\subsection{ Modelling of the Temperature Profile and the Effective Volumetric Factor}

The total molar content $\nu$ of the gas in the system related to the common pressure $p$:
\be \fr{\nu}{p}R=\{ V/T \}, \mbox{  where the effective ratio  }  \{V/T\} \ee
 is composed of:\\  
\begin{figure} 
\hspace{1cm}\includegraphics[scale=0.26]{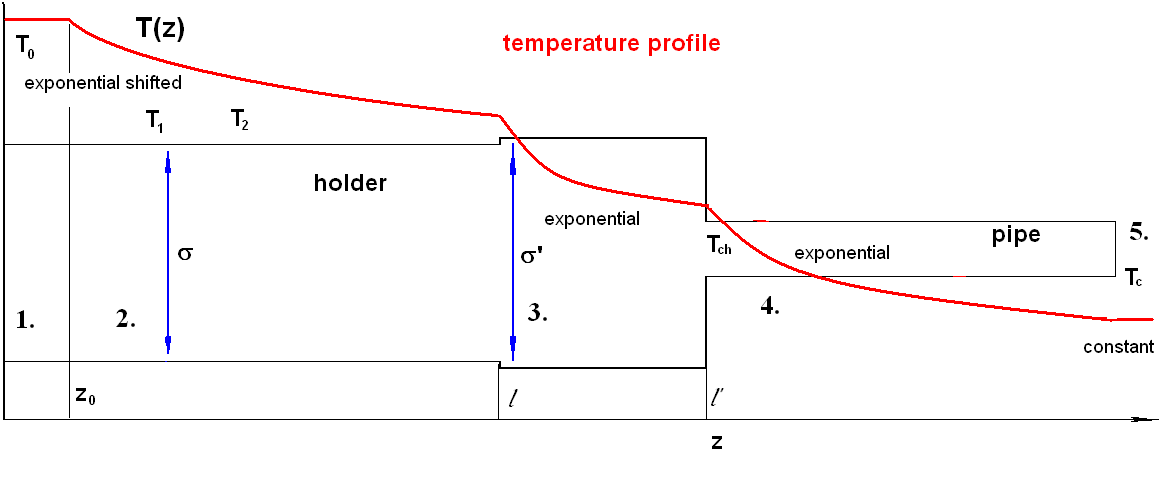}\\
\caption{Temperature profile of the volumetric gas}
\label{profile}
\end{figure} 
1. furnace heated part at $T_0$:  $ \{ V/T \}_1= z_0\si/T_0;$

2. first exponential part ( $ z_0\ra l$ ):
\be
\{ V/T \}_2=\int\limits_{z_0}^l \fr{dz}{T(z)} =
\fr{\si}{\bT\tau} \ln\left[1+ \fr{\bT}{T_0} \left( e^{\tau(l-z_0)}-1\right) \right]
\ee

3. second exponential part ( $l\ra l'$ ):
\be
\{ V/T \}_3=\int\limits_l^{l'} \fr{dz}{T(z)}=\si'\fr{l'-l}{\bT\left[\ln\fr{T_0-\bT}{T_{ch}-\bT}-\tau(l-z_0)
 \right]}
\cdot \ln\fr{T_{ch}}{(T_{ch}-\bT)\left[ 1+\fr{\bT}{T_0-\bT}e^{\tau(l-z_0)} \right] }
\ee

4. third exponential part (pipe of length L):
\be
\{ V/T \}_4=\int\limits_{pipe} \fr{dz}{T(z)}=\fr{V_p}{\bT \ln \fr{T_{ch}-\bT}{T_c-\bT}} 
\ln \left( \fr{\bT}{T_{ch}} \cdot \fr{T_{ch}-T_c}{T_c-\bT} + 1 \right)
\ee

5. cooled volumetric pool at $T_c$ corresponding to $25^oC:\ \   \{ V/T \}_5=V_s/T_c $\\

The effective ''reduced ratio'' $[V/T]$, that replaces the ratio $V/T$ for homogeneous temperature, reads:  
\be \{ V/T \}= \sum\limits_i \{ V/T \}_i \ee\\
 here:\\
$V_p=V_d-V_H$- the pipe volume,
$ \fr{\si'}{\si}=\left( \fr{r'}{r} \right)^2 $- ratio of two inner cross sections of the sample holder,\\
$V_H$- inner volume of the holder, $V_d$ - the measured 'dead volume': holder + pipe,\\
$V_s$- the system volume of the volumetric pool.\\ 
For practical purposes we will need also the effective ratio of ''dead volume'' $\{V/T\}_d={V/T}-V_s/T_c$:

\subsection{Verification of the Model}

  To construct the effective $V/T$ ratio, as proposed above, only the parameter $T_{ch}$, the temperature 
at the beginning of pipe channel, has not been measured. For 
practical calculation it was enough to evaluate this temperature roughly, since this channel is very thin 
and the final result is not sufficiently affected by some deviation from the exact value $T_{ch}$.
 
 For the initial state with "open valve", the gas in empty 
(without ad/desorbent) system has a homogeneous temperature distribution with $T=25^oC$ and the pressure
 $p_{25}$ = 197 kPa.
 If the furnace is set to hold the temperature of $450^oC$, 
the pressure grows up to $p_{450}$ = 225 kPa. 
 Since the gas amount remains the same, it is expected, that for effective ratios $\{V/T\}$ it holds:
\be \fr{ p_{25}}{p_{450}}= \fr{ \{V/T\}_{450} }{  \{V/T\}_{25}} \ee         
 
 Then we obtained the quotient $ p_{25}/p_{450}=0.8755 $. Supposed the $T_{ch}=150^oC$,
 (that seems to be quite realistic);
 The model predicts $ \{V/T\}_{450} = 9.799\cdot 10^{-8}$ and for temperature distribution of $ T= 25^oC$ it
provides $ \{V/T\}_{25} = 11.116\cdot 10^{-8}$, the quotient being required is 0.8815. 
 Therefore the agreement between calculated and the experimental data is fairly acceptable.   
 
\section{Conclusions}
In the current work it has been shown, that the measurement in a typical gas volumetric equipment
can be evaluated using the gas equation.
 The equation is modified by introducing the effective volumetric factor $\{V/T\}$ instead of
ratio $V/T$. This factor is calculated with help of the modelling of the temperature profile in the volumetric setup. 
 Using this procedure the effective volumetric factor could be obtained  for each heating temperature of the sample holder, as a proper parameter of the volumetric equipment. 
 A pre-calibration with a noble gas bevore a measurement is not necessary anymore.

\end{document}